\documentclass[prl,aps,10pt]{revtex4-1}
\usepackage{graphicx}
\usepackage{amsmath}
\usepackage{graphicx}

\begin{document}
\title{Baryon Annihilation and Regeneration in Heavy Ion Collisions}
\author{Yinghua Pan}
\affiliation{Department of Physics, Harbin Institute of Technology, Harbin, Heilongjiang 150006, People's Republic of China}
\affiliation{Department of Physics and Astronomy and National Superconducting Cyclotron Laboratory,
Michigan State University}
\author{Scott Pratt}
\affiliation{Department of Physics and Astronomy and National Superconducting Cyclotron Laboratory,
Michigan State University}
\date{\today}
\begin{abstract}
The role of baryon-antibaryon annihilation during the hadronic stage of a relativistic heavy ion collision is explored by simulating the chemical evolution of a hadron gas. Beginning with a chemically equilibrated gas at an initial temperature of 170 MeV, the chemical composition of a representative hydrodynamic cell is followed throughout the hadronic stage. The cell's volume changes with time according to a parameterization that mimics a three-dimensional hydrodynamic expansion. The chemical evolution includes both annihilation and regeneration of baryons, consistent with detailed balance. During the hadronic stage, the number of baryons drops by approximately 40\% for the case where there is no net baryonic charge.  When the calculations are performed without the baryon regenerating processes, e.g. $5\pi\rightarrow p\bar{p}$, the loss of baryons was found to be closer to 50\%. After accounting for annihilation, yields are consistent with measurements from the ALICE Collaboration at the LHC. Baryon annihilation is shown to alter the  extracted chemical breakup temperature by significantly changing the $p/\pi$ ratio. Assuming that annihilation cross sections are independent of the strangeness and isospin of the annihilating baryon and anti-baryon, the loss of strange baryons from annihilation is found to be similar.
\end{abstract}

\maketitle

\section{Introduction}

High energy heavy ion collisions provide a unique opportunity to investigate the properties of nuclear matter at high energy density, a few GeV/fm$^{3}$, which is several times the energy density of a proton. At these temperatures, $\gtrsim 170$ MeV, matter undergoes a transition between the hadronic phase and the quark-gluon plasma phase (QGP) \cite{Van:2002tm,McLarren:1986qg}. In the prevailing view, the strongly interacting matter validates the use of thermodynamics \cite{Stachel:1989th,Braun:1995th,Braun:1996th,Andronic:2011yq,Rafelski:1991th,Davidson:1991th,Sollfrank,Wheaton:2004qb} to describe the chemistry and hydrodynamics \cite{Kolb:2003dz} to describe the evolution. A typical assumption is that the hadronic chemistry is described by chemical freeze-out at a single temperature. This temperature is also associated with the QGP-hadron transition, or hadronization. Although hadrons certainly collide a few more times after hadronization, many models typically ignore the effects of hadronic-stage interactions on the chemistry. Particle ratios are often fit by a single temperature and a few chemical potentials to account for non-zero conserved charges such as strangeness, baryon number and electric charge. Additionally, fugacities have been added to account for the inability of the system to produce the requisite number of up, down and strange quarks for equilibrium \cite{Torrieri:2005va}. These pictures are usually based on the idea that, due to the strong interactions between quarks, all partitions into various hadron species are thermally sampled during hadronization, and that the evolution of chemical abundances after hadronization comes only from decays. However, the inferred temperature could be misleading if chemical reactions during the hadronization stage significantly change particle ratios.

Of course, the chemical make-up does not completely freeze-out --- hadrons indeed undergo chemical reactions during the hadronic stage. The effects of such evolution have been studied in two contexts. First, hadronic cascades (microscopic simulations that follow the trajectories of individual hadrons) such as URQMD \cite{URQMD} include inelastic processes.  Although these models include annihilation processes, e.g., $p\bar{p}\rightarrow n\pi$ \cite{Karpenko:2012yf,Steinheimer:2012bn,Steinheimer:2012rd,Becattini:2012sq,Becattini:2012xb}, few microscopic models have included the regeneration processes and therefore violate detailed balance. Inverse processes have been added to the hadronic HSD model, where $2\leftrightarrow 3$ processes such as $\rho\rho\pi\leftrightarrow p\bar{p}$ are included self consistently \cite{Cassing:2001ds}. Concomittantly, $2\leftrightarrow 3$ processes have also been incorporated into partonic simulations \cite{Xu:2004mz}. The particular process, $\bar{p}p\rightarrow \rho\rho\pi$, allows an annihilation to ultimately produce 5 mesons after the $\rho$ mesons decay. These incorporations into the HSD model have been used to investigate the possibility of creating the observed number of antibaryons through bottom-up processes, mostly at SPS, AGS and SIS energies. At higher energies, if one assumes that a chemically equilibrated number of baryons and antibaryons are created at hadronization, the issue becomes one of whether the regeneration process significantly dampens the annihilation rate during the hadronic stage. Implementing regeneration into cascade codes , as was done in HSD, requires significant investment, both in development and in the numerical cost of running more complicated codes. 

A second class of approaches involves solving for the chemical rates of a kinetically equilibrated gas. Given that there are hundreds of resonances, such models typically assign fugacities to a few numbers such as effective pion number, or the net number of baryons plus antibaryons, and then solve for the evolution of the fugacities \cite{Scott:1999ha,Chung:1997ch,Rapp:2002ha,Rapp:2000gy,Huovinen:2003sa}. In \cite{Rapp:2000gy} it was shown that regeneration rates, e.g. $n\pi\rightarrow p\bar{p}$, were important. Related issues were pursued by considering regeneration through Hagedorn states in \cite{NoronhaHostler:2007jf,Greiner:2004vm,NoronhaHostler:2009cf}, where it was estimated that chemical equilibration times were only fast enough to maintain equilibrium for temperatures $\gtrsim$ 170 MeV. Chemical evolution was superimposed onto a hydrodynamic calculation in \cite{Huovinen:2003sa},  where regeneration processes were included in a matter consistent with reproducing chemical equilibrium. In their results, annihilation and regeneration reduced baryon yields by 15-20\%. 

In all the calculations mentioned above, it is assumed that chemical equilibrium is maintained until some point, and that this temperature is sufficiently low that the system can be approximated as a hadron gas. This temperature, which we will label $T_0$, is sometimes associated with the hadronization temperature, though that is not necessary. Inferred temperatures near 170 MeV are common, even though the density of a hadron gas at such temperatures would suggest densities of one hadron per fm$^3$, which is roughly the inverse volume of a hadron. Further, lattice calculations show that the system is undergoing significant changes already by that time. The scalar quark condensate and the Polyakov loop are already significantly changed relative to vacuum values for temperatures $\sim 160$ MeV \cite{Bazavov:2013yv}. Fluctuations of conserved charges measured on the lattice show that hadron-like states exist up to somewhat higher temperatures, $\sim 200$ MeV \cite{Ratti:2011au}{Bazavov:2013dta}. We proceed here with calculations assuming $T_0=170$ MeV, even though hadrons at this temperature may have a rather different character than those in the vacuum.

The shape of pion, kaon and proton spectra can be well described by hybrid hydrodynamic/cascade models that assume chemical equilibrium at the hadronization temperature, and ignore any subsequent chemical evolution beyond resonance decays and reformation \cite{Song:2011hk,Novak:2013bqa}. However, such model's proton/pion ratio tends to significantly differ from experimental measurements from the PHENIX Collaboration at RHIC and from the ALICE Collaboration at the LHC. PHENIX ratios \cite{Adler:2003cb} were $\sim 35$\% below hybrid model predictions that fit the yields of pions and kaons in \cite{Song:2011hk,Novak:2013bqa} where baryon annihilations had been neglected. More consistent shortcomings are seen in Pb+Pb collisions at the LHC measured by ALICE \cite{Abelev:2012wca}. The $p/\pi$ ratio is nearly half of what one would expect from freezeout at hadronization temperatures \cite{Steinheimer:2012bn}, and are $\sim 50\%$ lower than the hydrodynamic results in \cite{Paatelainen:2012at}. Using a microscopic model, a reduction of approximately this size was seen \cite{Karpenko:2012yf} and in the erratum to \cite{Song:2011hk}, but those calculations ignore the inverse rates, which according to \cite{Rapp:2000gy} largely cancel the annihilation rates. In \cite{Steinheimer:2012rd}, the inverse process was crudely estimated at the level of 8\%, with this value then being used to represent the error associated with the lack of detailed balance in their model. The principal goal of this paper is to solve for the chemical rates of a wide range of hadronic species where baryon annihilation and regeneration are treated in a manner consistent with detailed balance, and to understand the relative importance of annihilation and regeneration. This is similar in spirit to what was outlined in \cite{Rapp:2000gy} and performed in \cite{Huovinen:2003sa}, but in more detail and with a larger number of resonances considered. The chemical evolution is also followed until interactions fully cease. We will determine whether in a consistent calculation annihilations might account for a $\sim 40-50$\% reduction in the final-state $p/\pi$ ratio measured in heavy-ion collisions at high energy relative to the final-state ration calculated without annihilation processes. Further, we discuss the importance of accounting for baryon annihilations when estimating the chemical freeze-out temperature from final-state yields. As a secondary goal, we wish to see whether strange baryons are affected at the same level.

In order to concentrate on chemistry, we employ a simplified model of the space-time evolution. We consider the evolution of a single hydrodynamic cell, where the volume changes as a function of time according to a simple prescription. The parameterization is based on the one-dimensional Bjorken expansion \cite{Bjorken:1983}, but is modified to account for transverse expansion. The parameterization is chosen to roughly match what happens in a three-dimensional hydrodynamic calculation of a central Pb+Pb collision at the LHC. In a more realistic model, one would consider the whole ensemble of hydrodynamic cells, each of which would evolve differently. Various species lose local kinetic equilibrium with one another when temperatures approach 140 MeV, as heavier species like protons cool faster than pions \cite{Pratt:1998gt}. Additionally, pions would begin to flow with different velocities, and leave the collision region while the embers would become relatively more baryon rich \cite{Sorge:1995pw}. Accounting for these non-equilibrium aspects requires microscopic simulations.  Nonetheless, this picture should be adequate to gauge the significance of annihilation and regeneration.

\section{Description of Calculation}

We model the hadronic stage by first assuming that all resonances are initially populated according to chemical equilibrium with a temperature $T_0$, set by default to 170 MeV, which is in the neighborhood of where  hadronization takes place. All chemical potentials are set to zero. The list of 319 resonances taken from the Particle Data Book \cite{particledatabook} extends to masses up to 2.25 GeV. Although one should have a non-zero baryonic chemical potential, even at high RHIC and LHC energies, it will be ignored for this study given that the our goal is to gauge the importance of annihilation effects.

After the initial time, we assume that collisions are sufficiently frequent to preserve local kinetic equilibrium, even though inelastic collisions are insufficient for maintaining chemical equilibrium. Each species is assigned a chemical potential, $\mu_i$. Each $\mu_i$ is initially set to zero, and if chemical equilibrium were maintained each $\mu_i$ would remain zero. Ignoring Bose effects, which are a $\sim 10\%$ correction for pions, thermal results within the context of the grand canonical ensemble for number densities, pressure, energy density and entropy are:
 \begin{eqnarray}\label{differentiation}
 n_{i}&=&\frac{g_i}{2\pi^2}
 \left(m_i^2TK_0(m_i/T)+2m_iT^2K_1(m_i/T)\right)e^{\mu_i/T},
 \nonumber \\
P&=&\sum_i n_iT,\\
\nonumber
\epsilon &=&\sum_i\frac{g_i}{2\pi^2}
\left[3m_i^2T^2K_0(m_i/T)+\left(m_i^3T+6m_iT^3\right)K_1(m_i/T)\right]e^{\mu_i/T},\\
\nonumber
s&=&\frac{P+\epsilon-\sum_i\mu_in_i}{T},
 \end{eqnarray}
where $g_{i}$ and $\mu_{i}$ are the degeneracy and chemical potential of the hadron species $i$, and $E_{i}=\sqrt{p^{2}+m_{i}^{2}}$ where $m_{i}$ is the particle mass and $K_{n}$ is the Bessel function of order $n$. 

For a chemically equilibrated system, the chemical potentials, $\mu_i$, are zero. For a system of massless particles in an isentropic expansion, the yields can be frozen and $\mu_i$ will remain zero. However, for massive particles the chemical potentials become positive if yields are frozen. This can be understood by considering the fact that the entropy per particle with zero chemical potential is near 3.5 for massless particles, but increases as a function of $m/T$. Even pions, the lightest mesons, develop significantly positive chemical potentials if the chemical evolution is frozen or if the evolution is confined to decays \cite{Greiner:1993jn,Scott:1999ha}. As will be seen later in this paper, chemical potentials remain positive even after decays, e.g. $\Delta\rightarrow N\pi$, and annihilations, e.g. $p\bar{p}\rightarrow 5\pi$ are included. Baryon annihilation reduces baryon chemical potentials, but again are insufficiently fast to maintain equilibrium. 

To return the pion chemical potential to zero, pion absorbing processes, e.g. $pp\pi_0\rightarrow p\Delta^+\rightarrow pp$, are required. Due to conservation of g-parity, there are only a few reactions in the meson sector that can reduce the final number of pions relative to the number one would get from decays alone. For example, $\rho+\pi\rightarrow\pi\pi$ violates g-parity. Reactions such as $\rho\rho\rightarrow\pi\pi$ or $\omega\pi\rightarrow\pi\pi$ conserve g-parity but are fairly rare. The baryon sector is a better candidate for affecting the final pion number since g-parity is not a constraint for reactions involving baryons. For instance $\Delta N\rightarrow NN$ lowers the final pion number by one. Such reactions might well reduce the final-state number of pions at the 5-10\% level. As will be shown here, pions become overpopulated by a factor $e^{\mu_\pi/T}\approx 1.75$ when the temperature has cooled to approximately 100 MeV. Thus, pion-absorbing reactions are far from able to maintain equilibrium, but they might also be non-negligible since chemical yields are now being evaluated at the 10\% level. Baryons comprise only $\approx 10\%$ of final-state hadrons at the LHC or at the highest RHIC energies, but represent a much higher fraction of the hadrons at SPS energies. Although such pion-absorbing processes are not considered here, they deserve further study, especially for analyses of lower-energy collisions.

Since we are interested in simply gauging the effects of annihilation, we consider a simplified picture of the expansion. In a one-dimensional boost-invariant expansion \cite{Bjorken:1983}, the volume of a fluid element would increase proportional to the time $\tau$. However, such a picture neglects transverse expansion. To crudely account for transverse expansion, we assume that the volume of the element increases as
\begin{equation}
\label{eq:OmegaOfTau}
\Omega(\tau)=\Omega(\tau_0)\frac{\tau}{\tau_0}\frac{\tau_\perp^2+\tau^2}{\tau_\perp^2+\tau_0^2},
\end{equation}
for times $\tau$ greater than the hadronization time $\tau_0$. By viewing the evolution of the density and temperature of hydrodynamic expansions from \cite{Vredevoogd:2012ui}, the parameters were chosen to be:
$\tau_0=8$ fm/$c$, $\tau_\perp=6.5$ fm/$c$. This parameterization produced density vs. time curves that were similar to those seen in full hydrodynamic models.


Based on the assumption above, the continuity equations can be easily be solved in Bjorken coordinates should all chemical rates be turned off,
 \begin{eqnarray}
n_{i}(\tau)=n_{i}(\tau_{0})\frac{\Omega(\tau_0)}{\Omega(\tau)}, \nonumber \\
s(\tau)= s(\tau_{0})\frac{\Omega)(\tau_0)}{\Omega(\tau)}.
\end{eqnarray}
The evolution equations for the chemistry are modified in the presence of chemical rates,
\begin{equation}
\label{eq:dndtau}
\frac{d}{d\tau}\left[\Omega(\tau) n_i(\tau)\right]=\Omega(\tau) R_i(\tau),
\end{equation}
where $R_i$ is net production/annihilation rate per unit $d^4x$ of particles of species $i$. The net rates, $R_i$, are all zero at equilibrium, but as the system cools the system loses equilibrium and the rates become non-zero (usually negative) to push toward equilibrium.

Once the system loses chemical equilibrium, entropy can be generated,
\begin{equation}
\label{eq:dsdtau}
\frac{d}{d\tau}\left[\Omega(\tau) s(\tau)\right]=-\Omega(\tau)\sum_i R_i(\tau)\frac{\mu_i(\tau)}{T(\tau)}.
\end{equation}
Given the entropy and number densities, one can solve for the temperature and chemical potentials. Thus, Given the rates $R_i$, Eq.s (\ref{eq:dndtau}-\ref{eq:dsdtau}) allow one to solve for the evolution of the number and entropy densities, or equivalently the chemical potentials and temperatures, as a function of $\tau$. A multidimensional Newton's method is used to numerically determine $\mu_i$ and $T$ from $n_i$ and $s$.

One contribution to $R_i$ is resonances decays. For the reaction $A\to a_{1}+a_{2}+\cdots+a_{n}$, the contribution to the growth rate for resonance of type $A$ is
 \begin{equation}
 R_A(\tau)=-\left(1- \exp\left\{\frac{\mu_{a_{1}}+\mu_{a_{2}}+\cdots+\mu_{a_n}-\mu_{A}}{T}\right\}\right)\left\langle\Gamma_{A\rightarrow a_1\cdots a_n}\right\rangle n_{A}(\tau).
 \end{equation}
Here $\langle\Gamma\rangle$ is the decay rate of the resonance, which is the nominal width divided by the average Lorentz time dilation factor,
\begin{equation}
\langle\Gamma\rangle=\frac{\Gamma}{(2\pi)^3n_A}\int(d^3p/E_A)~m_A f(p).
\end{equation}
Partial widths are taken from the particle data book \cite{particledatabook}. The contribution from the inverse reaction, $a_1+\cdots a_n\rightarrow A$, is accounted for by the factor $\exp[(-\mu_A+\mu_1\cdots +\mu_n)/T]$. This forces $R_A$ to go to zero when the chemical potentials balance. This channel also contributes to the growth rates for $a_1\cdots a_n$ with the same magnitude but opposite sign. All the hadronic decay channels from the particle data book are included in the calculation.

The second type of chemical reaction to be considered is that from baryon annihilation, $A+\bar{B}\rightarrow a_1+\cdots a_n$. The contribution to the growth rate of $A$ in this case is
\begin{equation}
\label{eq:annihilationdamping}
 R_A(\tau)=-\left(1- e^{(\mu_{a_{1}}+\mu_{a_{2}}+\cdots+\mu_{a_{n}}-\mu_A-\mu_B)/T}\right)
 \left\langle\sigma_{A+\bar{B}\rightarrow a_1\cdots a_n}v_{\rm rel}\right\rangle n_{A}(\tau)n_B(\tau).
\end{equation}
The average $\langle \sigma v_{\rm rel}\rangle$ for a relativistic gas is
\begin{equation}
\left\langle\sigma v_{\rm rel}\right\rangle=
\frac{1}{(2\pi)^6n_An_B}\int (d^3p_a/E_a)(d^3p_b/E_b)~f_a(p_a)f_b(p_b)\sigma(s)\sqrt{(p_a\cdot p_b)^2-m_a^2m_b^2}.
\end{equation}
We apply a simplified form for the $p\bar{p}$ annihilation cross section that is accurate for $P_{\rm lab}>100$ MeV/$c$ \cite{Wang:1998sh,ppbardata},
\begin{equation} \label{eq:xsection}
\sigma=67~P_{\rm lab}^{-0.7} {\rm mb}.
\end{equation}
Without particularly good justification, we further assume all baryon-antibaryon pairs have the same annihilation cross section as $p\bar{p}$. Averaged over momentum, the average $\langle \sigma v_{\rm rel}\rangle$ is near 5 fm$^2\cdot c$ for $p\bar{p}$. Other inelastic processes, e.g., $\Delta+N\rightarrow NN$, are ignored for this study as we are focusing on baryons. Such processes can change $n_\pi$ at the level of 10\% \cite{Scott:1999ha,bao:1995for}.

Experimental data show that the most likely number of pions in $p\bar{p}\rightarrow N\pi$ is $\sim 5$, and to simplify the calculation we assume that all annihilations proceed through the $5~\pi$ production channel. For pair annihilation with nonzero strangeness, we employ the minimal rule for kaon production, i.e, the number of kaons in the final state equals the initial strangeness. For example, if the initial strangeness is unity, then final production of annihilation will have one kaon. We still assume final production has five particles on average. The number of pions, $n_{\pi}$, will be $5-n_{K}$.

Before displaying numerical results, we summarize the assumptions in our calculation. We consider a system with initial charge set to zero, and at chemical equilibrium so that $\mu_i=0$ for all species. The initial temperature is set to 170 MeV, and the initial time is set to $\tau_0=6$ fm/$c$. The volume of the cell scales as in Eq. (\ref{eq:OmegaOfTau}). Only decays through the strong interaction are considered since both electromagnetic and weak decays are negligible on these time scales. The calculation considered 319 resonances, with masses up to 2.25 GeV/$c^2$. If experimental data or competing models include feed-down from weak decays, the model used here can easily be modified for the purposes of a consistent comparison.

\section{Results}

\begin{figure}[htbp]
  \centerline{\includegraphics[width=0.5\textwidth]{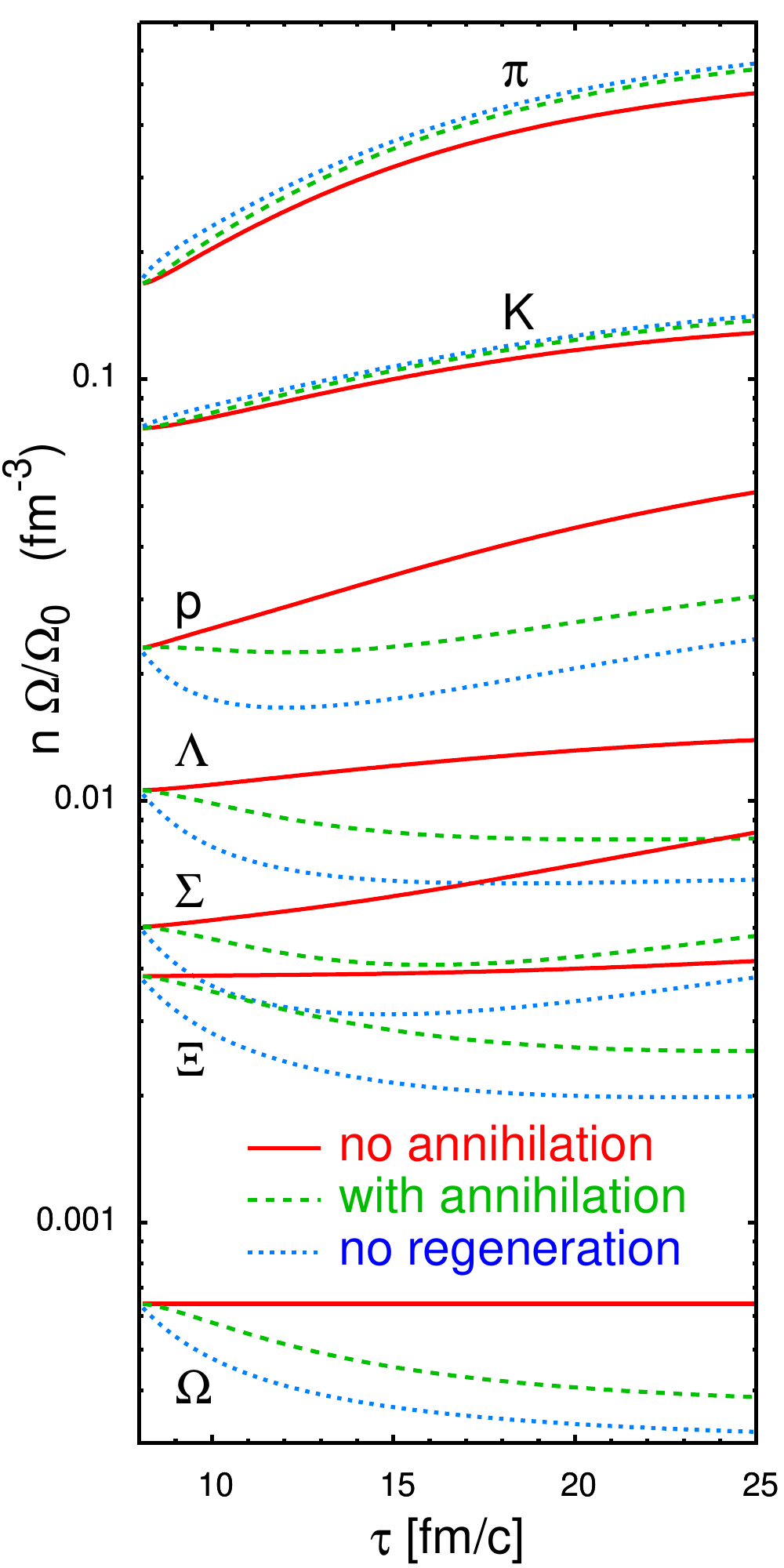}}
\caption{\label{fig:alldens} (color online)
Hadronic densitiies, scaled by the volume $\Omega(\tau)/\Omega(\tau_0)$, for the $\pi$ and $K$ mesons, and for the $p$, $\Lambda$, $\Sigma$, $\Xi$, and $\Omega$ baryons. In the absence of annihilation, the proton yield increases due to the decay of resonances like the $\Delta$. By adding annihilation, all baryon yields fall by $\sim 40$\%. Meson yields are modestly increased by the annihilation processes. If regeneration processes are ignored, the fraction of baryons that are annihilated increases to $\sim 50\%$
}
\label{Figure1}
\end{figure}

Figure \ref{fig:alldens} shows the evolution of the densities of pions, kaons, protons, Lambdas, Sigma, Cascade and Omega baryons. Results are displayed both for when annihilations are enabled and disabled. When  annihilations are disabled, the yields of stable particles increase due to feeding from the decays of unstable particles. Decays increase the number of protons by approximately 50\%, mainly through delta decays. The effect of baryon annihilation is to lower the baryon yields by $\sim$40\% relative to the yields without annihilation, to increase pion yields by $\sim$10\% and to increase the kaon yields by $\sim$5\%. Annihilations reduce the yields of strange baryons by approximately the same factor as protons. At $\tau_0$ annihilations play no role because the inverse process exactly cancels annihilation at equilibrium. However, within a few fm/$c$, the system is no longer at chemical equilibrium and the curves with and without annihilation diverge.

\begin{figure}[htbp]
  \centerline{\includegraphics[width=0.5\textwidth]{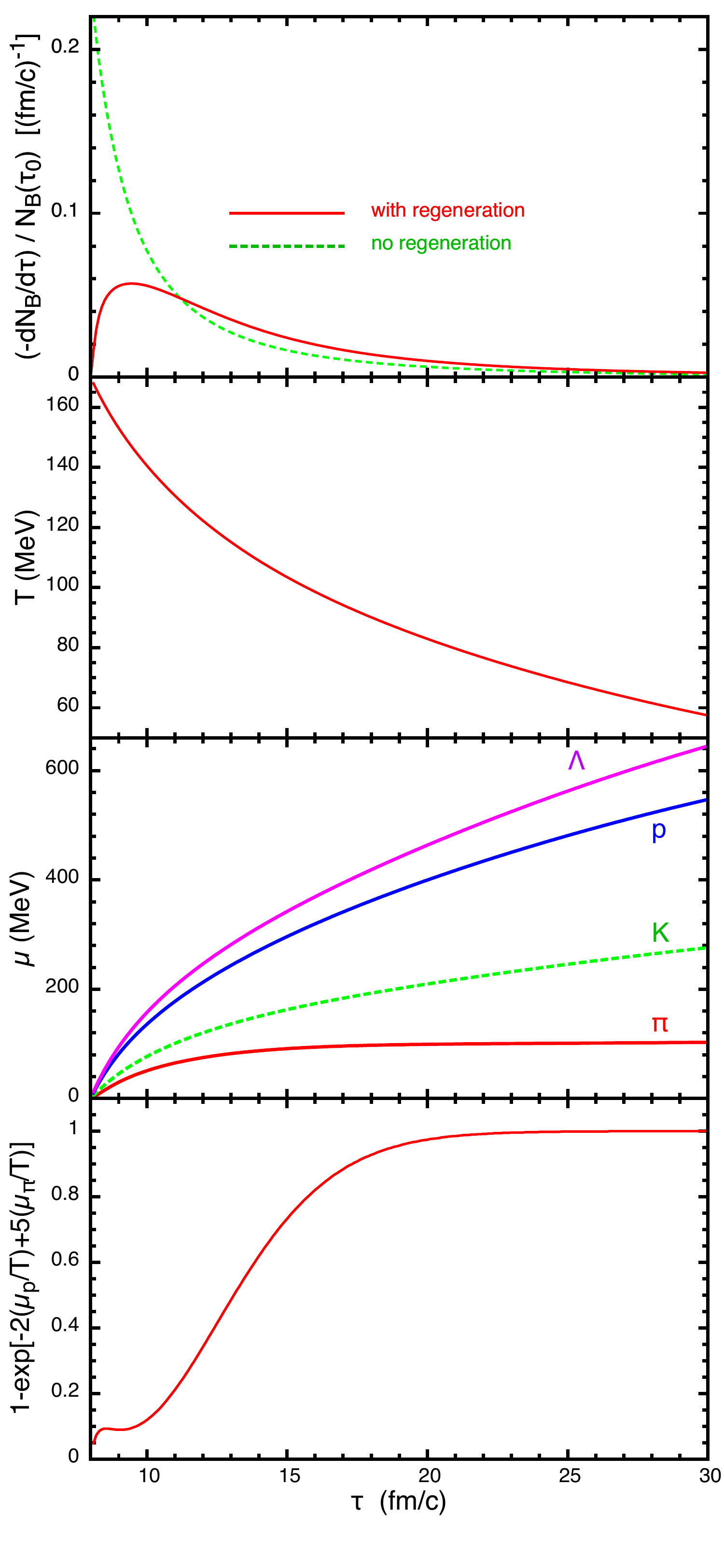}}
\caption{\label{fig:muT} (color online)
The annihilation rate (upper panel) as a function of time, beginning at the point when chemical equilibrium is lost, at $\tau_0=8$ fm/$c$. Regeneration blocks the annihilation for the first one or two fm/$c$, but after four or five fm/$c$ becomes negligible. When regeneration is included nearly half the annihilation occurs after the temperature has fallen below 125 MeV. The evolution of the temperature and chemical potentials for three species as a function of time in the middle two panels. The lower panel shows the fraction of the annihilation rate that is uncanceled by regeneration. In the latter stages, regeneration becomes negligible.
}
\end{figure}

As the system cools, chemical equilibrium is lost and the chemical potentials become non-zero. Figure \ref{fig:muT} shows the evolution of the temperature and chemical potentials as a function of time.  The chemical potentials develop more strongly for more massive particles since their yields are more sensitive to temperature. The lower panel demonstrates the degree to which the inverse processes cancel annihilation by considering the ratio $(1-e^{(-2\mu_p+5\mu_\pi)/T})$ which represents the fraction of annihilations canceled by regeneration in Eq(\ref{eq:annihilationdamping}). At the initial time, chemical populations are equilibrated and the inverse processes, e.g. $5\pi\rightarrow p\bar{p}$, exactly cancel the annihilation processes. However, the cancelation rapidly disappears, and by the end of the reaction, regeneration is negligible. From the upper panel one can see that annihilation continues for a significant time, and falls roughly proportional to the density, $\sim 1/\tau^3$. After accounting for regeneration, half the annihilation occurs after temperatures fall below 125 MeV. In \cite{Huovinen:2003sa} consistent chemical rates were used for a hydrodynamic calculation, but in that calculation freezeout was assumed for a temperature of 120 MeV. Such a choice, at least for LHC modeling, would miss a significant fraction of the annihilation. This may explain part of the reason why in \cite{Huovinen:2003sa} only 15-20\% of the baryons were annihilated, while in this calculation the rate was closer to 40\%.

\begin{figure}
\centerline{\includegraphics[width=0.6\textwidth]{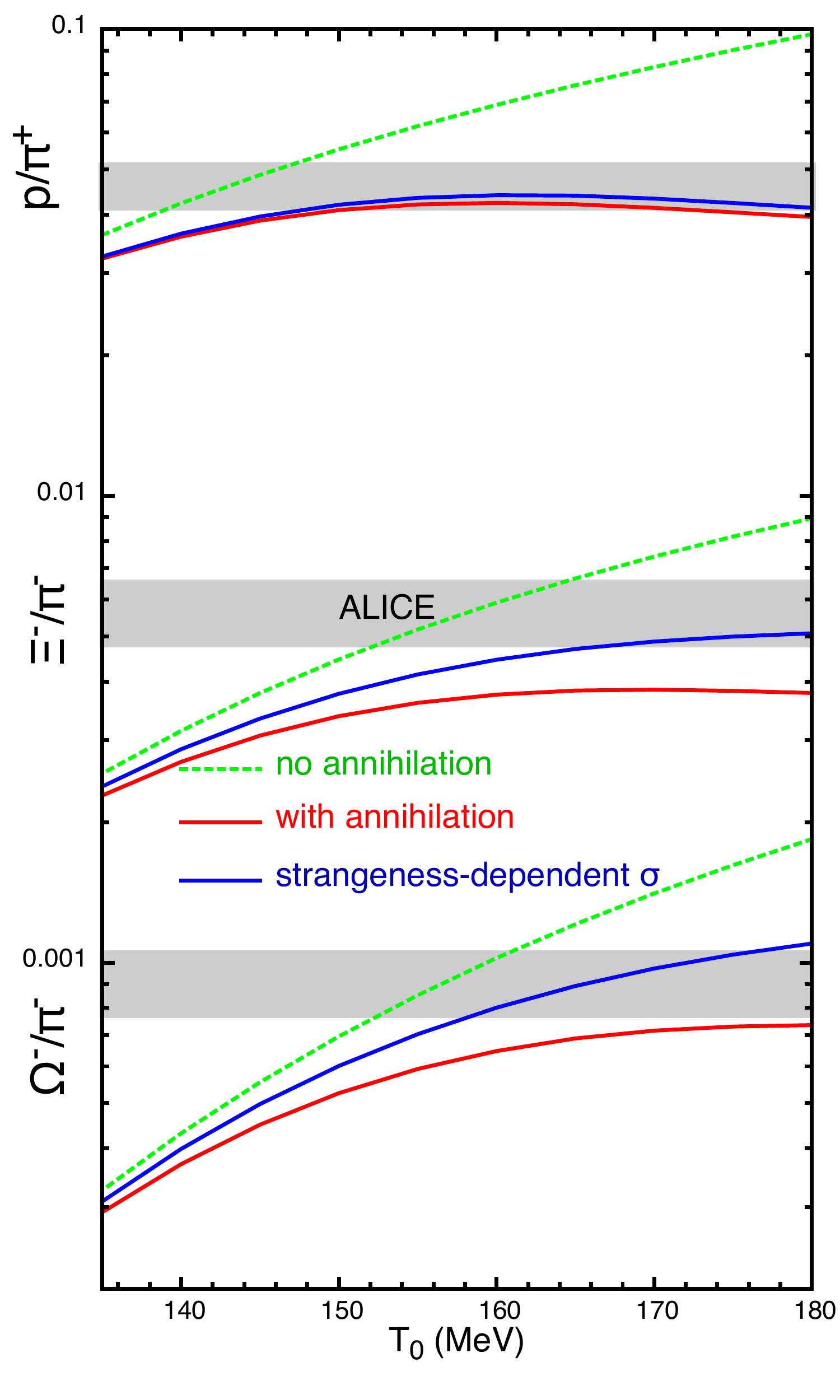}}
\caption{\label{fig:poverpi}(color online) 
The final $p/\pi^+$, $\Xi^-/\pi^-$ and $\Omega^-/\pi^-$ ratios as a function of $T_0$, the temperature at which chemical equilibrium is lost. Calculations with (solid red) and without baryon annihilation differ substantially. If one were to use the experimental $p/\pi^+$ ratio measured by ALICE to infer $T_0$, including annihilation would change the inferred value from $\sim 145$ MeV to a broad range of higher temperatures. If the annihilation cross sections were reduced for those baryons with a higher strangeness content, using the factor $\alpha=0.75$ in Eq. (\ref{eq:sreduction}), the resulting ratios are modestly higher for the multistrange baryons.
}
\end{figure}

A common use of particle ratios is to determine the chemical freeze-out temperature. Although analyses of this type involve a global fit, the temperature is mainly determined by ratios between the heaviest and lightest particles since they are the most sensitive to the temperature. The $p/\pi$ ratio is especially important. Since such analyses, c.g. \cite{Andronic:2011yq}, typically ignore baryon annihilation it is instructive to see the degree to which annihilation processes distort the extraction of a chemical freezeout temperature. Figure \ref{fig:poverpi} displays the final $p/\pi^+$ ratio as a function of the initial temperature for calculations with and without baryon annihilation. For initial temperatures different than 170, the initial time $\tau_0$ was adjusted so that the temperature vs time trajectory would closely match that of the default calculation with $T_0=$170 MeV and $\tau_0=8$ fm/$c$. In Fig. \ref{fig:poverpi} the calculations with annihilation also include regeneration. The ALICE collaboration at the LHC measured $p/\pi^+=0.46\pm 10\%$ \cite{Abelev:2012wca}, and is illustrated with a grey band in Fig. \ref{fig:poverpi}. Whereas this ratio would correspond to a chemical freeze-out temperature of 145 MeV if baryon annihilation were ignored, if one accounts for the effects of baryon annihilation and regeneration, one becomes fairly insensitive to $T_0$, and any values greater than $\gtrsim 150$ MeV seem acceptable.

Figure \ref{fig:poverpi} also shows two hyperon to pion ratios, $\Xi/\pi$ and $\Omega/\pi$. Using the same cross section for the annihilation of any two baryon species, independent of the isospin, spin, or strangeness content of the annihilating baryons, the annihilation fraction turned out to be largely independent of strangeness. Of course, there is little reason to think that the annihilation cross section for $\Omega-\bar{\Xi}$ should be the same as for $p\bar{p}$. It is known that $\bar{p}n$ cross sections are similar to $\bar{p}p$ \cite{pbarn}, so one might feel justified in assuming that annihilation only weakly depends on isospin. However, virtually nothing is known about how it might be affected by strangeness. There are model-dependent calculations \cite{kovacs} or \cite{Kapusta:2002pg}, which are unconstrained by data. For example, in \cite{Kapusta:2002pg} cross sections are calculated in a simple field theory incorporating SU(3) flavor symmetry, and the chemical relaxation times for various hyperons vary on the order of a few tens of percent from that for protons. But the assumptions for the form of the field theory are difficult to justify for the large time-like momenta transfer involved. Further, one needs cross sections for excited baryons states, such as $\Delta$s and $\Omega$. The annihilation cross section for hyperons are also lower according to the additive quark model \cite{URQMD}. It is safe to state that annihilation cross sections could easily vary by several tens of percent, and that these variations would be very difficult to calculate from first principles. One can also use fits to heavy ion data, but in these cases one needs to first rely on an assumption for the initial production rates in dense matter. 

One might expect annihilation cross sections to be smaller for hyperons since hyperons have smaller charge radii than nucleons. For that reason, we consider a simple parameterization where the dependence of the annihilation cross section on strangeness is encapsulated in a single parameter, $\alpha$. We then repeated our calculations above with different values of $\alpha$. The form for the cross section for the annihilation of two baryons $A$ and $B$ with relative three-momentum $P$ is
\begin{equation}
\label{eq:sreduction}
\sigma_{\bar{A}B}(Q)=\sigma_{\bar{p}p}(Q)\left\{\frac{1}{2}
\left[\alpha^{|S_A|}+\alpha^{|S_B|}\right]\right\}^2.
\end{equation}
This expression describes a physical picture where the particles annihilate whenever the distance of closest approach is less than the sum of the two baryon's radii $R_A$ and $R_B$, and where the size of the baryon scales as
\begin{equation}
R_A=R_p\alpha^{|S_A|}.
\end{equation}
For $\alpha=1$, one reproduces the cross sections described in Eq. (\ref{eq:xsection}). For $\Omega\bar{\Omega}$ annihilation the cross section is reduced by a factor $\alpha^3$. Results for $\alpha=0.75$ are displayed in Fig. \ref{fig:poverpi}. Even though this is a rather strong dependence on the strangeness content, the change in the annihilation fractions are rather modest. Ratios for $\Xi/\pi$ and $\Omega/\pi$ ratios are also shown from ALICE \cite{Abelev:2013zaa}. The ALICE results are in the range of the predictions of the model that include baryon annihilation. 

The chemical rate calculations ignore $2\leftrightarrow 2$ processes that affect the hyperon/proton ratios. For example, $p+K\leftrightarrow \Lambda +\pi$ has a preference for moving strangeness from kaons into hyperons due to the fact that $\mu_p+\mu_K>\mu_\Lambda+\mu_\pi$. This would increase the $\Lambda/p$ ratio. In contrast, the process $K+\Lambda\leftrightarrow p+\pi$ prefers to reduce the number of strange quarks and reduce the $\Lambda/p$ ratio. Since these processes involve momentum transfers on the order of a few hundred MeV, they can probably be calculated with some confidence. In the URQMD calculations of \cite{Steinheimer:2012rd,Becattini:2012sq,Becattini:2012xb} the number of $\Omega$s and $\Lambda$s increased in the hadronic stage, which is in contrast to the results found here. It would be useful to further analyze the URQMD evolution to understand what is driving this difference. 

\section{Conclusions}

Figure \ref{fig:alldens} shows that annihilation processes during the hadronic stage play a significant role in the baryo-chemistry of the quark-gluon plasma. Calculated baryon yields that include the effects of annihilation are reduced by $\sim 40\%$ for central Pb+Pb collisions at the LHC. This helps explain the magnitude of the $p$ and $\bar{p}$ spectra at the LHC relative to the pion spectra without surrendering the assumption of chemical equilibrium at the advent of the hadronic stage. Figure \ref{fig:alldens} also displays results from calculations where the inverse reactions, e.g. $5\pi\rightarrow p\bar{p}$, were not included. In this case baryon yields would have fallen approximately 50\% if regeneration had not been incorporated. This is consistent with microscopic transport models that also ignore regeneration \cite{Steinheimer:2012rd,Becattini:2012sq,Becattini:2012xb}. Regeneration increases this reduced yield by about 20\% so that the net reduction in baryon number is close to 40\%. This is in line with observations at the LHC, which have uncertainties at the 15\% level. Although it is clearly important to include annihilation, one may or may not wish to include regeneration depending on the level of accuracy required for a specific analysis. Certainly, $n\rightarrow 2$ processes can require significant costs, both in the development and the execution of the code. Regeneration perfectly cancels the annihilation rate immediately after chemical equilibrium, then becomes negligible by the end of the reaction. This implies that baryo-chemistry is most sensitive to the physics for temperatures $\lesssim 140$ MeV. For these temperatures, hadronic cascades are well justified. Given that the assumption of local kinetic equilibrium becomes increasingly questionable once the temperature falls below 140 MeV, the results found here are suspicious. However, the annihilation fraction without regeneration was compared to that of a microscopic simulation and found to match within one or two percent. Certainly, the effects of regeneration could vary a few percent between this simple model and a more realistic model. Thus, this simple calculation mainly serves as a means to gauge whether regeneration, which requires significant work to implement in a microscopic model, warrants the effort to  incorporate into microscopic models. The answer is ``yes'', if one wants to understand baryon yields to the 10\% level or better, but ``no'' if 20\% accuracy is sufficient. It is also clear that baryon annihilation cannot be ignored even if 20\% accuracy is sufficient. Even after regeneration is included, the annihilation fraction is in the neighborhood of 40\%.

Figure \ref{fig:alldens} also showed the effect of annihilation on strange baryon yields. However, these results are not trustworthy given the lack of knowledge of annihilation cross sections for hyperons. Whereas the nucleon-antinucleon cross sections do not strongly depend on isospin, e.g. $\bar{p}p$ and $\bar{p}n$ annihilation cross sections are similar \cite{pbarn}, little is known about the annihilation cross sections for hyperons, either with antinucleons or antihyperons, and one must resort to unconstrained model calculations \cite{kovacs} or to fits to particle yields in heavy ion collisions \cite{Wang:1998sh}. Additionally, it is possible that $2\rightarrow 2$ inelastic processes such as $K+p\rightarrow \pi+\Lambda$, which are not included here, might affect yields. In our calculations $\mu_K+\mu_p>\mu_\Lambda+\mu_\pi$, so such effects might alter the $\Lambda/p$ ratio. Any study of hyperon yields in heavy ion collisions will remain largely speculative until experimental information is obtained regarding hyperon annihilation reactions in the vacuum.

The analysis of this paper, and of similar works, would significantly benefit if the details of how yields were measured was clearly provided. This is especially true for understanding the degree to which weak resonance decays are included in experimental results, especially preliminary results. For example, at the 10\% level ratios of baryon yields to pions depend on whether weak decays, e.g. $K_s\rightarrow \pi\pi$, that produce pions are included. Presentations of model results are often similarly vague in their description. All weak decays were consistently ignored in the calculations presented here, but that can easily be modified to better either match experimental conditions or to match the choices of a competing theoretical model.

Although sufficient for obtaining an understanding of the importance of annihilation chemistry during the hadronic stage, the model employed here is too simplistic to compare to data at better than the 10\% level. The assumptions are numerous: hadronic equilibrium at the onset of hadronization, boost-invariant dynamics, and local kinetic equilibrium that persists for long times. Further, some of the chemistry is neglected. Inelastic collisions, such as $\Delta N\rightarrow NN$, can lower the final meson yields on the order of 10\%. As shown here, to make a hadronic cascade reliable to the 10\% level, one should employ a fully consistent microscopic simulation of the hadronic stage that includes regeneration. The cascade would then account for the three-dimensional expansion, include more $2\leftrightarrow 2$ processes, provide a realistic freeze-out picture, and allow for the loss of local kinetic equilibrium.

\end{document}